\documentclass[aip,amsmath,amssymb,reprint]{revtex4-1}

\usepackage{graphicx}
\usepackage{bm}
\usepackage[utf8]{inputenc}
\usepackage[x11names]{xcolor}
\usepackage[T1]{fontenc}
\usepackage{siunitx}
\usepackage{gensymb}
\usepackage{sidecap}
\usepackage{ulem}

\newcommand{\mumm}{\micro\meter\squared}

\begin{document}


\title{\centering Optimal architecture for diamond-based wide-field thermal imaging}

\author{R.~Tanos}
\affiliation{Laboratoire Charles Coulomb, Universit\'{e} de Montpellier and CNRS, Place Eug\`ene Bataillon, 34095 Montpellier, France}
\author{W.~Akhtar}
 \affiliation{ Department of Physics, JMI, Central University, New Delhi, India}
\author{S.~Monneret}
 \affiliation{Institut Fresnel, CNRS, Aix Marseille Univ, Centrale Marseille, Marseille, France}
\author{F.~Favaro de Oliveira}
\affiliation{Qnami, Klingelbergstrasse 82, CH-4056, Basel, Swizterland}
\affiliation{Department of Physics, University of Basel, Klingelbergstrasse 82, Basel CH-4056, Switzerland}
\author{G. Seniutinas}
\affiliation{Qnami, Klingelbergstrasse 82, CH-4056, Basel, Swizterland}
\author{M.~Munsch}
\affiliation{Qnami, Klingelbergstrasse 82, CH-4056, Basel, Swizterland}
\author{P.~Maletinsky}
\affiliation{Qnami, Klingelbergstrasse 82, CH-4056, Basel, Swizterland}
\affiliation{Department of Physics, University of Basel, Klingelbergstrasse 82, Basel CH-4056, Switzerland}
\author{L.~le~Gratiet}
\affiliation{Centre de Nanosciences et de Nanotechnologies, C2N, CNRS -  Universit\'e Paris-Sud -  Universit\'e Paris-Saclay, 10 Boulevard Thomas Gobert, Palaiseau, France}
\author{ I.~Sagnes}
\affiliation{Centre de Nanosciences et de Nanotechnologies, C2N, CNRS -  Universit\'e Paris-Sud -  Universit\'e Paris-Saclay, 10 Boulevard Thomas Gobert, Palaiseau, France}
\author{A.~Dr\'eau}
\affiliation{Laboratoire Charles Coulomb, Universit\'{e} de Montpellier and CNRS, Place Eug\`ene Bataillon, 34095 Montpellier, France}
\author{C.~Gergely}
\affiliation{Laboratoire Charles Coulomb, Universit\'{e} de Montpellier and CNRS, Place Eug\`ene Bataillon, 34095 Montpellier, France}
\author{V.~Jacques}
\affiliation{Laboratoire Charles Coulomb, Universit\'{e} de Montpellier and CNRS, Place Eug\`ene Bataillon, 34095 Montpellier, France}
\author{G.~Baffou}
 \affiliation{Institut Fresnel, CNRS, Aix Marseille Univ, Centrale Marseille, Marseille, France}
\author{I.~Robert-Philip}
\affiliation{Laboratoire Charles Coulomb, Universit\'{e} de Montpellier and CNRS, Place Eug\`ene Bataillon, 34095 Montpellier, France}

\begin{abstract}
Nitrogen-Vacancy centers in diamond possess an electronic spin resonance that strongly depends on temperature, which makes them efficient temperature sensor with a sensitivity down to a few mK/$\sqrt{\rm Hz}$. However, the high thermal conductivity of the host diamond may strongly damp any temperature variations, leading to invasive measurements when probing local temperature distributions. In view of determining possible and optimal configurations for diamond-based wide-field thermal imaging, we here investigate, both experimentally and numerically, the effect of the presence of diamond on microscale temperature distributions. Three geometrical configurations are studied: a bulk diamond substrate, a thin diamond layer bonded on quartz and diamond nanoparticles dispersed on quartz. We show that the use of bulk diamond substrates for thermal imaging is highly invasive, in the sense that it prevents any substantial temperature increase. Conversely, thin diamond layers partly solve this issue and could provide a possible alternative for microscale thermal imaging.  Dispersions of diamond nanoparticles throughout the sample appear as the most relevant approach as they do not affect the temperature distribution, although NV centers in nanodiamonds yield lower temperature sensitivities compared to bulk diamond.

\end{abstract}
\pacs{}
\maketitle

Thermal imaging, enabling fast and accurate monitoring of heat distribution at sub-micron scales, has become decisive in a broad range of fields from exploratory research up to prototyping and manufacturing in nanomaterials science, nanoelectronics, nanophotonics or nanochemistry. Various detection schemes are being explored in this respect \cite{ReviewThermometry}. These schemes include tip-enhanced infrared or Raman thermometry~\cite{Weng, Li}, scanning thermal microscopy (SThM) \cite{Majundar, Gomes}, SQUID-based nano-thermometry~\cite{HalbertalNature2016} or nanoscale fluorescence thermometry making use of  fluorescent nanoparticles either dispersed on the probed sample or attached to the tip of an atomic force microscope (AFM)~\cite{Aigouy}. Yet none of these techniques can simultaneously provide fast, sensitive (in the sub-K/$\sqrt{\rm Hz}$ range), and quantitative thermal imaging with a sub-micron spatial resolution under ambient conditions. 
\begin{figure}[t]
\begin{centering}
\includegraphics[width=88 mm]{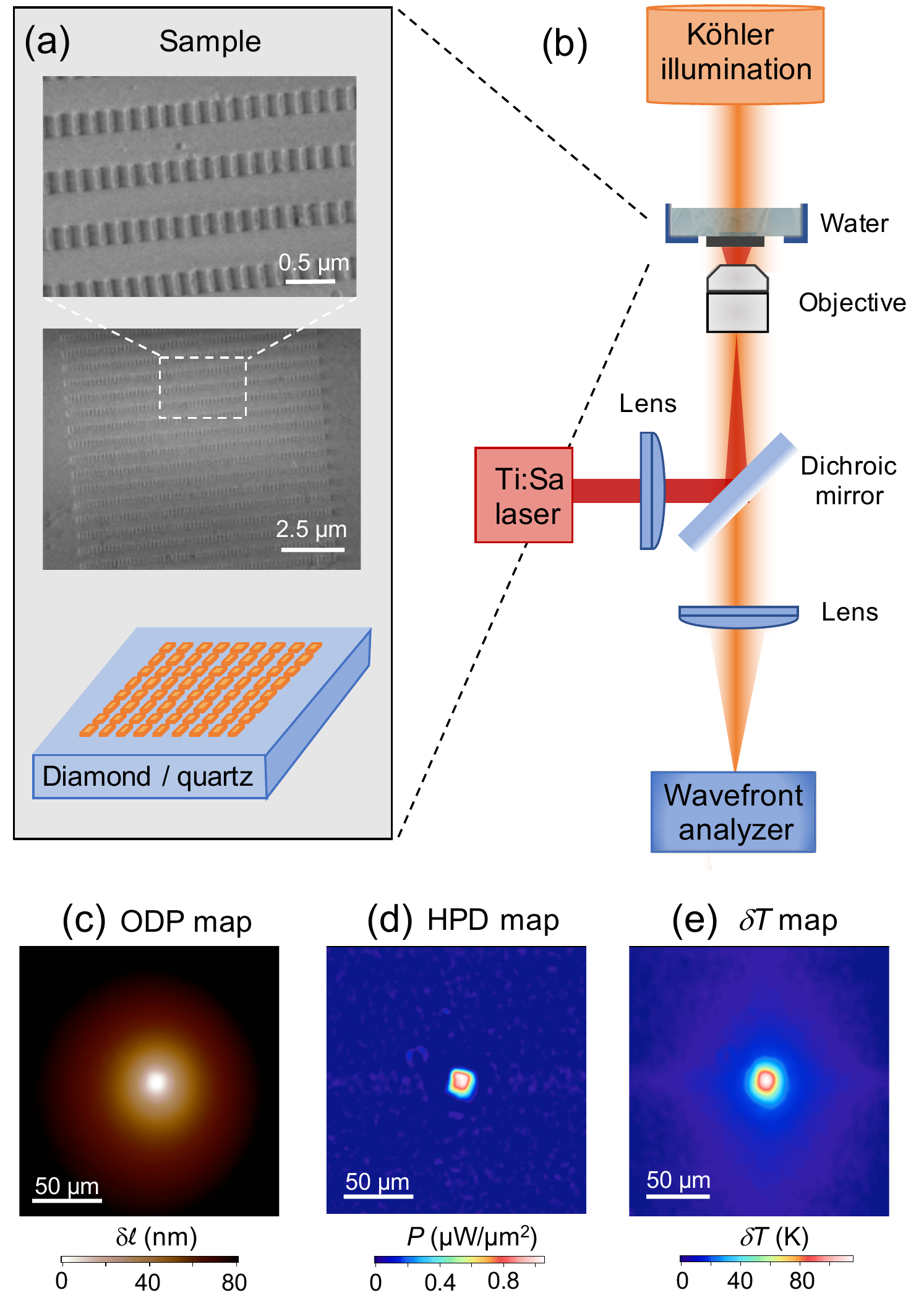} 
\end{centering}
\caption{(a) Scanning electron images of a thermoplasmonic structure deposited on quartz. (b) Schematic of the thermal imaging experimental setup. The thermoplasmonic arrays are  surrounded by water and illuminated in a wide field configuration through a high numerical aperture microscope objective (NA=0.95) by a tunable infrared Ti:Sapphire laser whose wavelength is set close to the cuboids plasmonic resonance. A spatially coherent white-light source impinges on the sample on the opposite side, passes through the same objective and is focused on a wavefront analyzer, which consists of a modified Hartmann grating combined with a CCD camera (Sid4Bio, Phasics SA). The wavefront analyzer records an interferogram on the CCD, that is further processed to retrieve the optical path difference (OPD) generated by the thermal-induced refractive index variations in water. (c) Typical OPD map recorded for a thermoplasmonic array consisting of $60\times 120$~nm$^2$ cross-section cuboids illuminated at $760$~nm on the quartz substrate. (d),(e) Resulting maps of heat power density (HPD) and temperature variations.}
\label{Fig1} 
\end{figure}

Nitrogen-Vacancy (NV) centers in diamond have garnered growing attention in the last decade, notably because their electron spin resonance can be detected optically~\cite{Gruber1997} and strongly depends on various external perturbations. This dependence has enabled to implement highly sensitive NV-based quantum sensors capable to locally probe several physical quantities including strain~\cite{Trusheim_2016,TetienneNanoLett2019}, electric~\cite{Dolde2011,doi:10.1021/acs.nanolett.9b00900} and magnetic fields~\cite{Degen,Rondin,QDM}. The sensing capabilities of the NV center have also been extended to thermometry~\cite{KucskoNature}, building on the variation of the zero-field splitting parameter of its electron spin sublevels with temperature~\cite{Acosta}. A thermal sensitivity in the range of $100$~mK/$\sqrt{\rm Hz}$ was demonstrated for single NV centers hosted in nanodiamonds~\cite{Neumann2013} and can reach values down to few~mK/$\sqrt{\rm Hz}$ while using NV centers  with long spin coherence times embedded in ultrapure bulk diamond samples~\cite{KucskoNature,Wang2015,Neumann2013,Toyli8417}.  Besides building on the temperature dependance of the spin resonance, temperature sensing with NV centers in diamond, can utilize the temperature dependance of their optical response (Debye-Waller factor, zero-phonon line or Anti-Stokes photoluminescence wavelength)\cite{Traneaav9180,Plakhotnik}; such all-optical schemes have also been extended to other point defects in diamond, including SiV, GeV or SnV colored centers \cite{Nguyen2018,Fan2018,Alkahtani}. Last but not least, the temperature sensitivities of NV centers can be pushed down to few tens $\mu$K/$\sqrt{\rm Hz}$, by relying on a transduction of temperature gradients on magnetic fields probed by their electronic spins\cite{PhysRevX.8.011042,Liu2019}.

These thermal sensing modalities can be exploited for thermal imaging. A first strategy consists in grafting a NV-doped nanodiamond at the apex of either an optical fiber or an AFM tip, which is scanned above the surface of the sample to be probed~\cite{Fedotov,Laraoui,Tetienne}. The spatial resolution of thermal imaging is then ultimately limited by the nanodiamond size. Such a {\it scanning}-NV configuration however requires a complex experimental apparatus and suffers from long acquisition times. In addition, the thermal sensitivity is impaired by the short spin coherence time of NV centers hosted in nanodiamonds. Another approach makes use of {\it stationary} NV centers  in a wide-field detection scheme~\cite{Neumann2013,Wang2015,Andrich}. Although the spatial resolution is then fundamentally limited by diffraction ($\sim 500$~nm), this approach offers faster acquisition times and a simple experimental configuration. To date, wide-field thermal imaging has been solely demonstrated with NV-doped nanodiamonds directly dispersed on the surface of a target sample. 

In this paper, we investigate whether this method could be extended to NV centers hosted in an ultrapure bulk diamond sample in view of improving the thermal sensitivity thanks to the improved NV spin coherence time. To this end, we first experimentally analyze the temperature distribution induced by a micron-size heat source deposited on the surface of a bulk diamond material. The results are then compared to numerical simulations performed for a bulk diamond, a $100-$nm thin diamond membrane deposited on quartz, and nanodiamonds directly dispersed on the heat source. This study shows that any detection scheme based on bulk diamond is highly perturbative due to the large thermal conductivity of diamond. Dispersions of nanodiamonds in the vicinity of the heat source, appear as the most relevant configuration for invasive-less wide-field thermal imaging, while thin diamond membranes could provide an effective option for thermal imaging at microscale.  

The heat source here consists of an array of gold nanoparticles illuminated at their plasmonic resonance wavelength~\cite{baffou2013thermo, baffou2013thermoarray}. We consider $10\times\SI{10}{\mumm}$  assemblies of gold cuboids with a thickness $t=40$~nm, a width $w=60$~nm, and a length $l$ ranging from $90$~nm to $210$~nm. As illustrated in Figure~\ref{Fig1}(a), the metallic cuboids are arranged in periodic two-dimensional rectangular patterns, each with constant cuboid dimensions and with a side-to-side distance of $100$~nm along the cuboid's width direction and of $1,6\times l$ along the cuboid's length direction. These thermoplasmonic arrays were fabricated on two substrates, namely an ultrapure electronic grade diamond substrate grown by chemical vapor deposition (Element 6) and on a quartz substrate used as reference. The process combines an electron-beam lithography step, the deposition of 5 nm of titanium and 35 nm of gold, followed by a lift-off. Under continuous near-infrared laser illumination at their absorption resonance, the thermoplasmonic structures give rise to a delocalised temperature distribution, fairly uniform throughout the array~\cite{Richardson, baffou2013thermoarray}. The increase in temperature induced by the array is then governed, inter alia, by the absorption cross-section $\sigma_{\rm abs}$ of each cuboids, the number $N$ of cuboids forming the array and the intensity $\mathcal{I}$ of the infrared laser illumination.

Temperature variations induced by the thermoplasmonic structures were imaged using quadriwave lateral shearing interferometry \cite{TIQSI} [see Fig.~\ref{Fig1}(b)]. This technique uses a wavefront analyser to indirectly  provide temperature distributions with a diffraction-limited spatial resolution and an accuracy of about $1$K. It relies on the temperature-dependent refractive index of a surrounding water layer covering the thermoplasmonic arrays.  The heat generated by the arrays results in a steady-state temperature distribution, which provokes modifications of the refractive index in the water. These spatial variations in refractive index result in a wavefront distortion of a white light source illuminating the sample in a K\"ohler configuration. Such local distortions are recorded on a wavefront analyser, providing spatially-resolved optical path difference (OPD) maps. A typical OPD map recorded for a thermoplasmonic array deposited on quartz is shown in Figure~\ref{Fig1}(c). Post-processing of these data then enables retrieving the spatial distribution of the heat power density (HPD) [Fig.~\ref{Fig1}(d)] and of the spatial temperature variations $\delta T$ [Fig.~\ref{Fig1}(e)]. This processing relies on a modeling of heat diffusion based on Green's functions formalism~\cite{TIQSI}. 

\begin{figure}[t]
\includegraphics[width=90 mm]{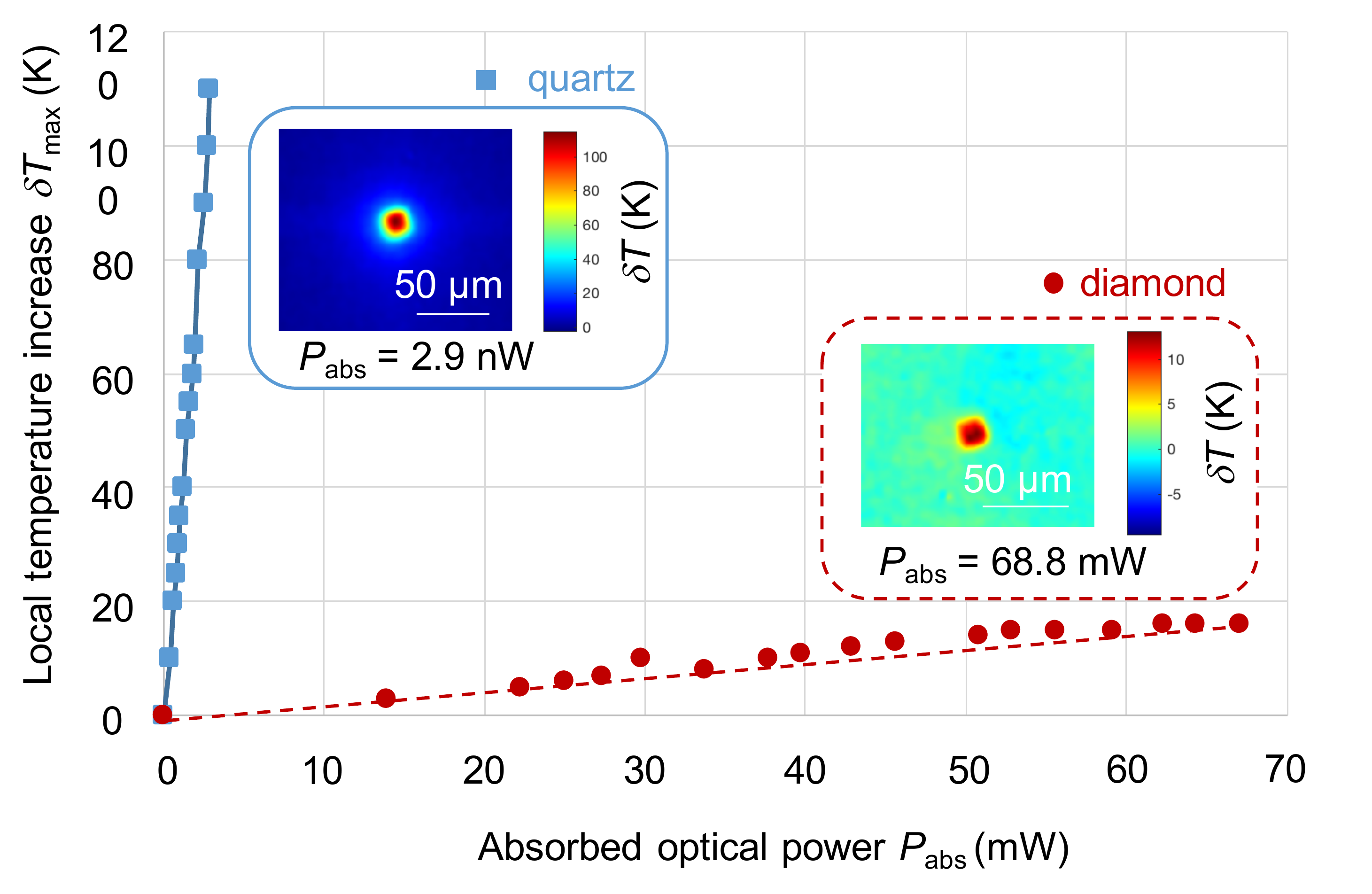}
\caption{Maximum local heating $\delta T_{\rm max}$ as a function of the absorbed optical power $P_{\rm abs}$ for thermoplasmonic arrays deposited on quartz (blue squares) and diamond (red dots). Insets: Temperature maps recorded (i) on the quartz substrate for $P_{\rm abs}=2.9$~mW leading $\delta T_{\rm max}=115.2^\circ$C (top within the blue frame) and (ii) on the diamond substrate for an absorbed laser power of $P_{\rm abs}=68.8$~mW leading $\delta T_{\rm max}=12.8^\circ$C (bottom within the red frame). Solid lines are linear fits of the experimental data.}
\label{Fig2} 
\end{figure}

The plasmonic resonance wavelength of the heating structures varies both with the substrate and the shape of the metallic cuboids. In order to obtain a reliable comparison between the heat generated on an ultrapure diamond substrate and on a quartz substrate, experimental conditions leading to similar absorption cross-sections of the thermoplasmonic arrays were carefully selected. As explained in details in Ref.~\cite{TIQSI}, the absorbed power $P_{\rm abs}$ can be inferred through a spatial integration of the HPD image. The absorption cross-section is then given by $\sigma_{\rm abs}=P_{\rm abs}/(\mathcal{I}\times N)$, where $\mathcal{I}$ is the infrared laser intensity (power per unit area)  and $N$ the number of cuboids forming the thermoplasmonic array. Similar absorption cross-sections $\sigma_{\rm abs}\sim 2\times10{^4}$~nm$^2$ were found for (i) $60\times120$~nm$^2$ cross-section cuboids illuminated at $760$~nm on the quartz substrate and for (ii) $60\times 90$~nm$^2$ cross-section cuboids illuminated at $710$~nm on the diamond substrate. 

With such experimental conditions, thermal images were recorded while increasing the absorbed infrared laser power $P_{\rm abs}$. As shown in Figure~\ref{Fig2}, the maximum local heating $\delta T_{\rm max}$ evolves linearly with $P_{\rm abs}$ for both substrates, as expected. However, for equivalent absorption powers, the heating induced by the thermoplasmonic array on the quartz substrate is much higher than the one observed on diamond. A difference of two orders of magnitude is observed, with a maximum heating per absorbed power of $38$~K/mW and $0.2$~K/mW for quartz and diamond substrates, respectively. This discrepancy results from the large difference in thermal conductivity $\kappa$ of the substrates. While $\kappa$ is about $1.4$~W/(m.K) in quartz, it reaches values almost three-order of magnitudes higher in diamond with $\kappa=1000-3300$ W/(m.K) depending on its purity. Consequently, while heat dissipation in quartz material is rather low, heat transfer in diamond is highly efficient. As such, the diamond substrate acts as a thermal sink that cools the heat source \cite{Satura}. Any architecture of thermal imagers based on bulk diamond will thus be perturbative: the measured temperature, though recorded with a high sensitivity, will be the temperature of the heat source however significantly cooled by the diamond substrate hosting the NV centers probes, down to temperatures that may be too small to be measured. This precludes the use of bulk diamond samples for wide-field thermal imaging. 

\begin{figure}[b]
\includegraphics[width=90 mm]{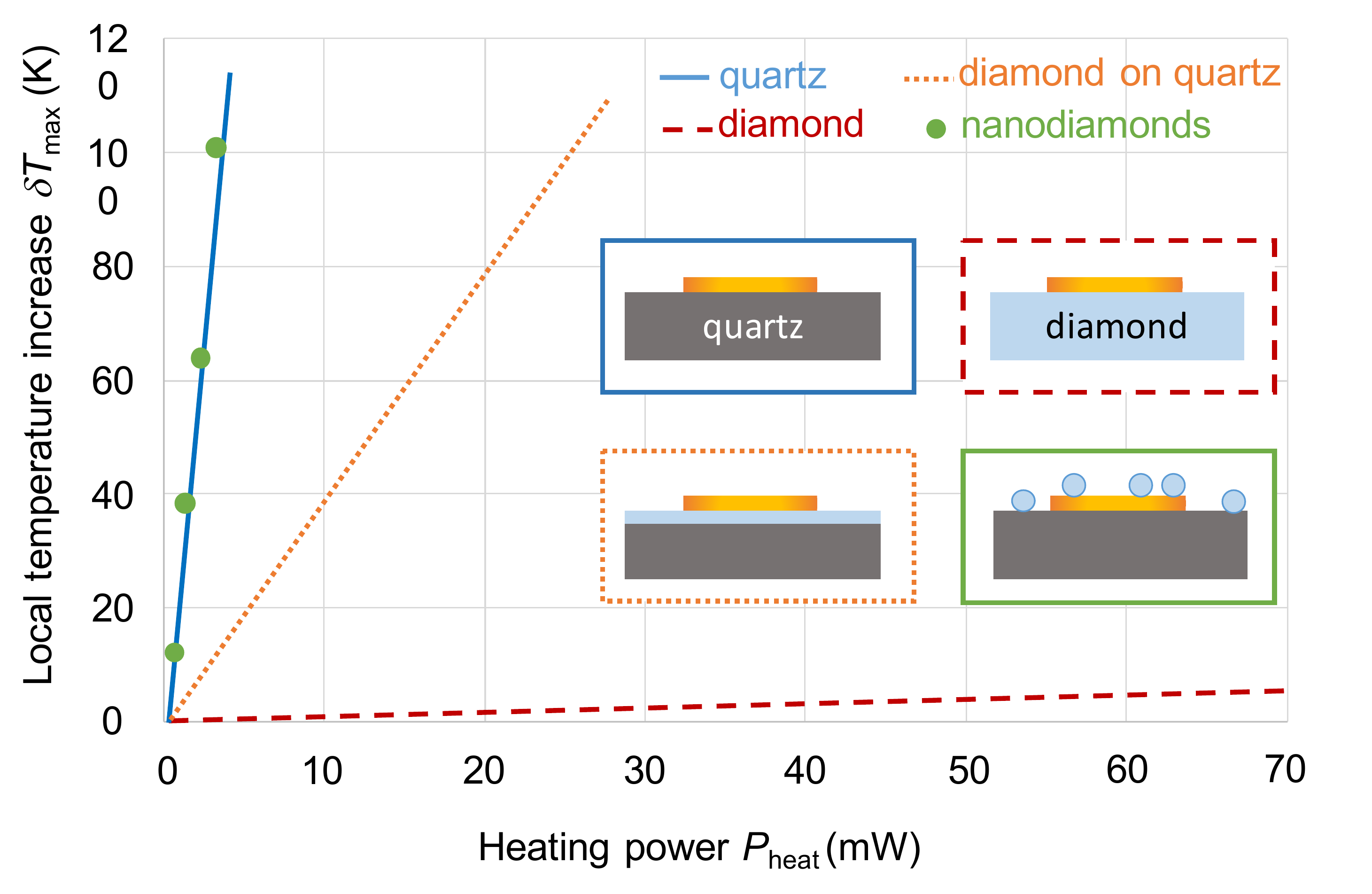} 
\caption{Numerical simulations of the maximal temperature heating $\delta T_{\rm max}$ at the sample surface induced by a two-dimensional $10\times\SI{10}{\mumm}$ heater for increasing heat powers  $P_{\rm heat}$. Four configurations are considered: the hot plate is located either on a diamond substrate (red dash line),  on a quartz substrate (blue solid line), on a $100$-nm thick diamond membrane bonded on a quartz substrate (orange dotted line) or on a quartz substrate with diamond nanostructures dispersed on the surface (green dots). In this latter architecture, the temperature increase is the one reached within the nanostructures whose dimensions are in the few tens of nanometers range. The insets show schematics of the four configurations.  The  temperature increase per heating power reaches  $\simeq4$~K/mW in the heterostructure geometry; it is  almost two orders of magnitude higher than the one predicted on bulk diamond and is solely $\simeq6,7$ times smaller than the one expected on the quartz geometry.}
\label{Fig3} 
\end{figure}

In order to reduce heat dissipation in bulk diamond, one alternative approach could be to rely on a thin diamond membrane bonded on a low thermal conductivity substrate, such as quartz. Figure~\ref{Fig3} gathers finite-difference time-domain (FDTD) simulations based on heat conduction modeling of the temperature increase as a function of the heat power for three different diamond-based architectures: (i) a bulk diamond, (ii) a $100-$nm thin diamond membrane on quartz, and (iii) nanodiamonds dispersed on the heat source. The achieved temperature increase in the diamond-quartz heterostructure, is much higher than the one predicted for bulk diamond, and gets closer to the one expected on quartz.  This stems from the strongly reduced effective thermal conductivity of the diamond-quartz heterostructure, in which the effective heat release through diamond occurs no longer in three but in two dimensions. In view of applications for thermal imaging, this partly lifts the limitations encountered with bulk diamond. However the heat diffusion throughout the diamond layer would still yield a lateral spreading of the temperature distribution over a length scale of few $\mu$m's, thus limiting the spatial resolution. 

Conversely, the temperature of nanodiamonds exactly matches the one expected on the low thermal conductivity quartz substrate (see Fig.~\ref{Fig3}), with a homogenous temperature distribution across the entire nanodiamond volume. This thermalization of the nanodiamonds with the local thermal bath, combined with the constant temperature profile within the nanoparticle, allows for a non-perturbative measurement of the heat source temperature, despite the randomness in the position of the NV spin sensor inside the nanodiamond. 

These results enable us to specify the optimal configuration for NV-based wide-field thermal imaging. Though bulk diamond imagers would provide the highest temperature sensitivity due to increased spin coherence times, efficient heat transfer in such high thermal conductivity material unavoidably lead to a significant cooling of the sample to be probed and thus prevents any effective temperature mapping at sub-micron scales. Engineered heterostructures stacking thin diamond membranes over low conductive substrates, feature significantly reduced heat diffusion, providing a valuable option for imaging at microscale. The most relevant configuration for non-invasive wide-field thermal imaging yet consists of random or ordered dispersions of diamond nanostructures on the hot sample to be probed. These thermal imagers may bring competitive imaging modalities with implications in the fields of micro- and nano-electronics, nanoplasmonics, or chemistry at nanoscale.\\

The authors acknowledge financial support from the French Research National Agency (through ANR THESEUS grant), from the center National de la Recherche Scientifique (through  CNRS TEMPO grant) and from the European Commission (through H2020 program's ASTERIQS, TSAR and ThermoQuant grants). This work was also partly supported by the french RENATECH network. \\

\bibliography{Biblio}

\end{document}